\documentclass{caosp}

\usepackage{graphicx}
\articleNo{182} \pubyear{2009} \volume{39} \volnumber{39} \firstpage{18} \lastpage{24}
\received{November 17, 2008}
\accepted{January 26, 2009}

\begin{document}
%\htitle{Release of meteoroids ...}
\hauthor{L.\,Korno\v{s}, J.\,T\'{o}th and P.\,Vere\v{s}}
%\hauthor{L.\,Korno\v{s} {\it et al.}}

\title{Release of meteoroids from asteroids by Earth's tides}

\author{
        L.\,Korno\v{s}
      \and
        J.\,T\'{o}th
      \and
        P.\,Vere\v{s}
       }

\institute{Department of Astronomy, Physics of the Earth and Meteorology,
Faculty of Mathematics, Physics, and Informatics, Comenius University, 842 48
Bratislava, The Slovak Republic}

%           \email{kornos@fmph.uniba.sk}

\date{November 8, 2008}

\maketitle

\begin{abstract}
The orbital evolution of particles released from the surface of a rubble-pile
body by Earth's tides during flyby within the Roche limit is studied. Test
particles initially placed on the surface leave the surface and escape the
parent body. Released particles remain in a relative small cloud for about 500
years and spread evenly along the orbit of the parent asteroid during next
several hundred years. Their orbital elements exhibit very small dispersion in
the mentioned time frame.
\end{abstract}

\section{Introduction}

\label{intr}

The motivation for this work was the fall of the Neuschwanstein meteorite on
April 6, 2002 observed by the European Fireball Network. The analysis of
photographic records has shown (Spurn\'{y} {\it et al.}, 2003) that the
heliocentric orbit of the object is practically identical to the orbit of the
well-known P\v{r}\'{i}bram meteorite, which was photographed 43 years earlier.
The different meteoritic types, P\v{r}\'{i}bram being an H5 ordinary chondrite
(Ceplecha, 1961) with cosmic-ray exposure age 12 Myr (Stauffer and Urey, 1962)
and Neuschwanstein an EL6 enstatite chondrite with cosmic-ray exposure age 48
Myr (Bischoff and Zipfel, 2003; Zipfel {\it et al.}, 2003), make their common
origin very problematic. Spurn\'{y} {\it et al.} (2003) suggested the existence
of a heterogeneous stream in the P\v{r}\'{i}bram orbit. Moreover, there is
evidence of material movement near the surface of the well-known asteroid
Itokawa (Miyamoto {\it et al.}, 2007\,a, 2007\,b). In this way, boulders on the
surface of the parent asteroid are differentially exposed by cosmic rays. This
effect might be responsible for different cosmic-ray exposure ages of potential
meteoroids in an asteroidal stream.

Korno\v{s} {\it et al.} (2008) have shown a similar orbital evolution of both
meteorites at least for 5000 years in the past, which supports the possible
existence of a meteoroid stream in the orbit of P\v{r}\'{i}bram. Furthermore,
the observation of the fireball "Malacky" with the orbit similar to
P\v{r}\'{i}bram (Spurn\'{y}, 2008) increases the expectancy of meteoroid stream
existence. On the other hand, the origin of such a stream is still an open
question. Pauls and Gladman (2005) showed that the occurrence of pairs as close
as P\v{r}\'{i}bram and Neuschwanstein is consistent with a random probability.

The idea of meteoroid showers associated with asteroids was proposed in the
past, first time suggested by Olivier (1925) and Hoffmeister (1937). The
existence of asteroidal-meteoritic streams was published by Halliday {\it et
al.} (1990). Generally, several mechanisms of meteoroids' release from
asteroids were suggested and one of them could be the Earth's tides during
close NEA approaches. This scenario offers low relative velocities of
meteoroids with respect to the parent asteroid and, at the same time, their
orbits are close to the Earth's orbit. Thus the potential meteor shower might
be observable from the Earth. The frequency of such approaches (inside the
Roche limit of the solid Earth body $\sim 2 R_{\otimes}$) by potential parent
objects of size of $\sim$ 300 meters is one per 25\,000 years according to the
population distribution of NEAs by Ivanov (2006). Pauls and Gladman (2005)
showed that the decoherence time of possible streams in orbits of well-known
meteorite falls (Innisfree, Peekskill, and P\v{r}\'{i}bram) is about 50\,000
years and more depending on the orbit type. Thus, by Pauls and Gladman, an
extremely recent breakup of the parent body would be required.

In this paper, we model the orbital evolution of particles released from the
surface of an Itokawa-like asteroid (a rubble-pile body significantly covered
by unbound pebbles and rocks) by Earth's tides during the asteroid flyby within
the Roche limit.

\section{Model}

The complex models of tidal disruption of an asteroid were presented in several
papers, e.g. Bottke {\it et al.} (1997, 1998), Richardson {\it et al.} (1998),
Sharma {\it et al.} (2006), Holsapple and Michel (2008). Following mentioned
results a fragile rubble pile asteroid with weak internal cohesion starts to
deform into a rotational ellipsoid shape when getting close to the Roche limit
of the Earth. Afterwards the differential gravitational tidal force from the
Earth overcomes the gravity of the asteroid and its surface layers at the ends
of the longest axis start to separate from the body. Even the most part of the
asteroid survives the close approach at the cost of resurfacing and shape
elongation, it easily loses a significant number of unbound surface or
near-surface pebbles, rocks and boulders.

Our analysis incorporates the parent body of meteoroids as a rubble-pile
asteroid with no tensile strength and with the mass and the longest axis same
as for asteroid Itokawa (Abe {\it et al.}, 2006). According to previously
mentioned papers we assume that our model asteroid will undergo through tidal
elongation up to 700\,m in the longest axis.

We gradually put 100 test particles onto the surface of the asteroid in the
plane of motion while the asteroid was moving inside the Roche limit. In the
constant time intervals a pair of particles was set in the line of centers of
the Earth and the asteroid, giving two particles on the opposite sites of the
asteroid. Then each particle is traced from its initial distance of 350\,m from
the center of the asteroid. In the first stage we examined the motion of the
particle in the geocentric coordinates of Earth -- asteroid -- particle in a
classical three-body problem (the Earth, the asteroid and a particle of
negligible mass). The motion of the asteroid was approximated by hyperbolic
motion with respect to the center of the Earth and the numerical integration
solved the motion of each particle from its entry to the Roche limit to the
100\,000 km distance from the center of the Earth. When particles reached the
100\,000 km distance the second stage started. We transformed the position
vectors and velocity vectors into the cartesian heliocentric ecliptic
coordinate system and the further motion was computed by direct integration of
particles' equations of motion including gravitational influences of all
planets. The Earth and the Moon were considered as two independent objects. The
integration was done for 1000 years to the future.

\section{Results}

When analyzing the motion of test particles released by tides from the asteroid
we used two types of orbits of the parent body that cross the Earth's Roche
limit: the orbit of meteorite P\v{r}\'{i}bram ($a$=2.40\,AU, $e$=0.67,
$i$=10.5$\degr$, Apollo type) which had to be slightly modified in the mean
anomaly to miss the center of the Earth in about 13\,000 km. Secondly, we
analysed the orbit of asteroid 2004\,FU162 ($a$=0.83\,AU, $e$=0.4,
$i$=4.1$\degr$, Aten type) that passed 0.000086\,AU (= 13\,000\,km) from the
center of the Earth on March 31.65, 2004 ({\it MPEC 2004-Q22}). This body has
the absolute magnitude ($H$=28.7) corresponding to 5-10\,m in diameter. Given
its small size, we assume this body is a monolith and therefore it could not be
a source of meteoroids.

\begin{table}[h]
\small
\begin{center}
\caption{The maximum escape velocity and relative distance of the particle from
the parent asteroid when the asteroid reaches the 100\,000\,km distance from
the Earth. Parent asteroid disintegration is taken into account (whole or half
of the parent body mass survives).} \label{t1}
\begin{tabular}{l|ll|ll}
\hline\hline
mass & 1.0 & &0.5& \\
\hline
orbit type & escape velocity & distance & escape velocity & distance\\
\hline
2004\,FU162 & 10\,cm s$^{-1}$ & 795\,m & 11\,cm\,s$^{-1}$ & 975\,m \\

P\v{r}\'{i}bram & 6\,cm\,s$^{-1}$ & 430\,m & 7\,cm\,s$^{-1}$ & 550\,m \\
\hline\hline
\end{tabular}
\end{center}
\end{table}

%Fig. 1
\begin {figure}
\centerline{\includegraphics[width=3.4cm,angle=-90]{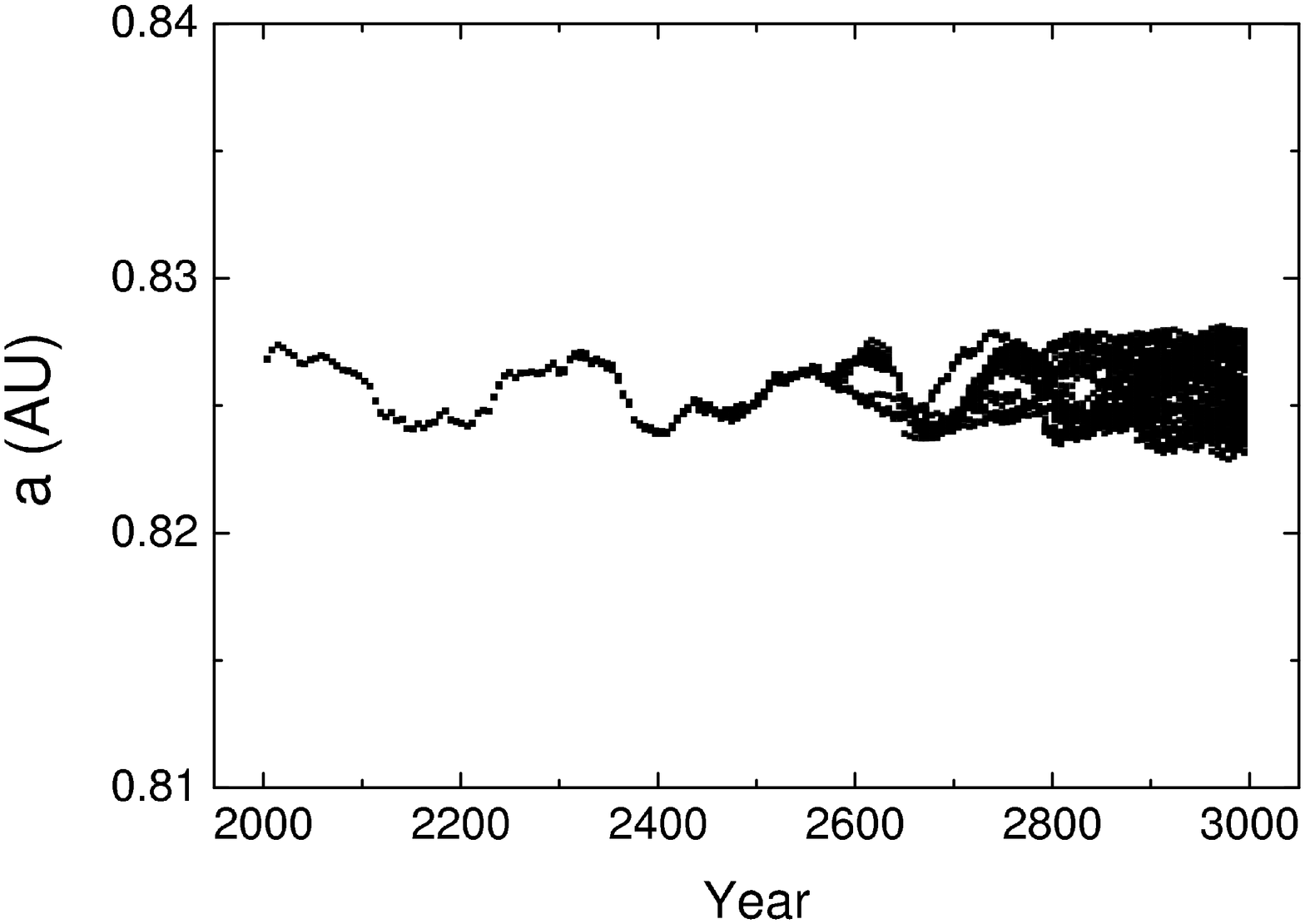}
            \hspace{0.6cm}
            \includegraphics[width=3.4cm,angle=-90]{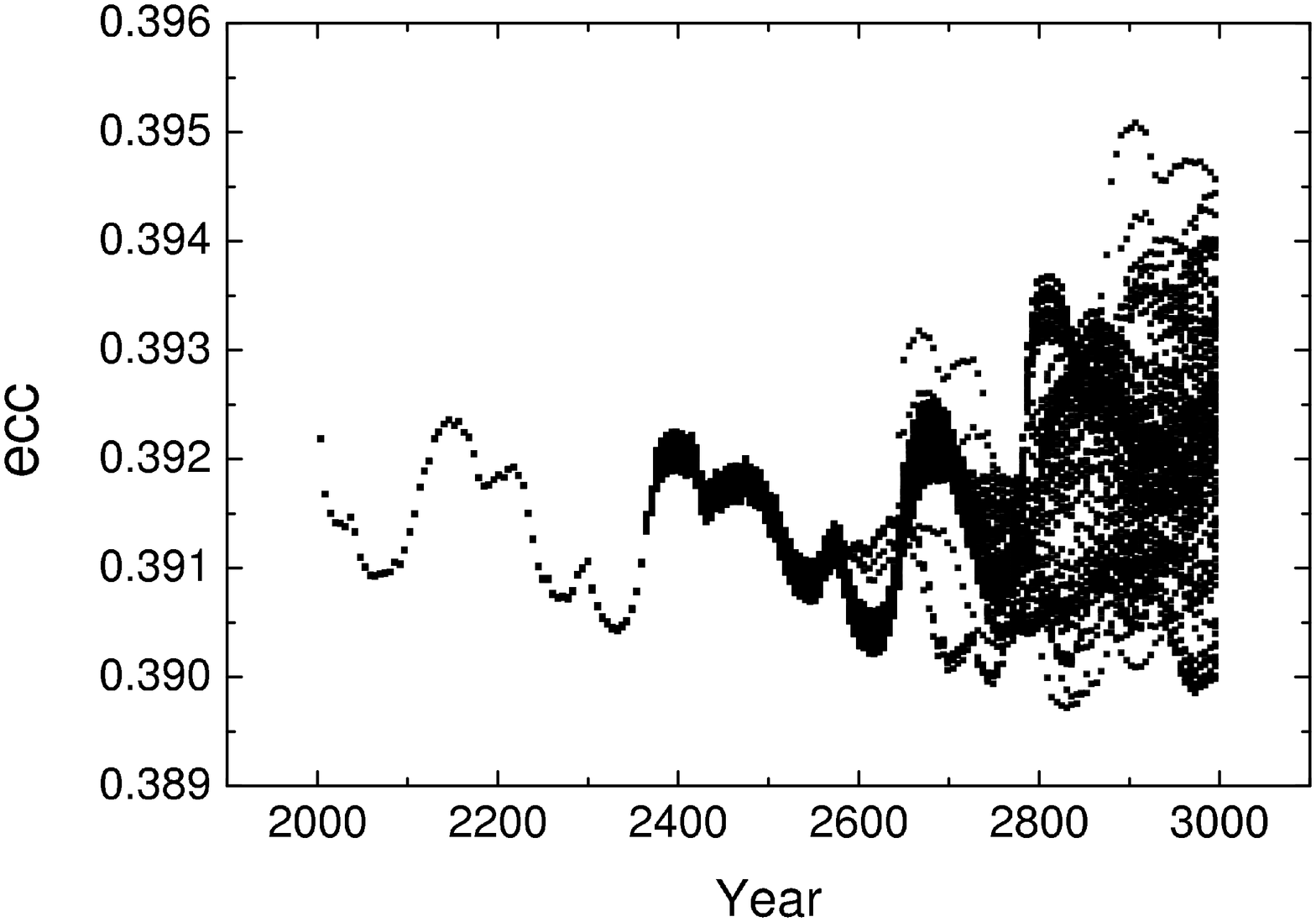}}
\vspace{0.4cm} \centerline{\includegraphics[width=3.4cm,angle=-90]{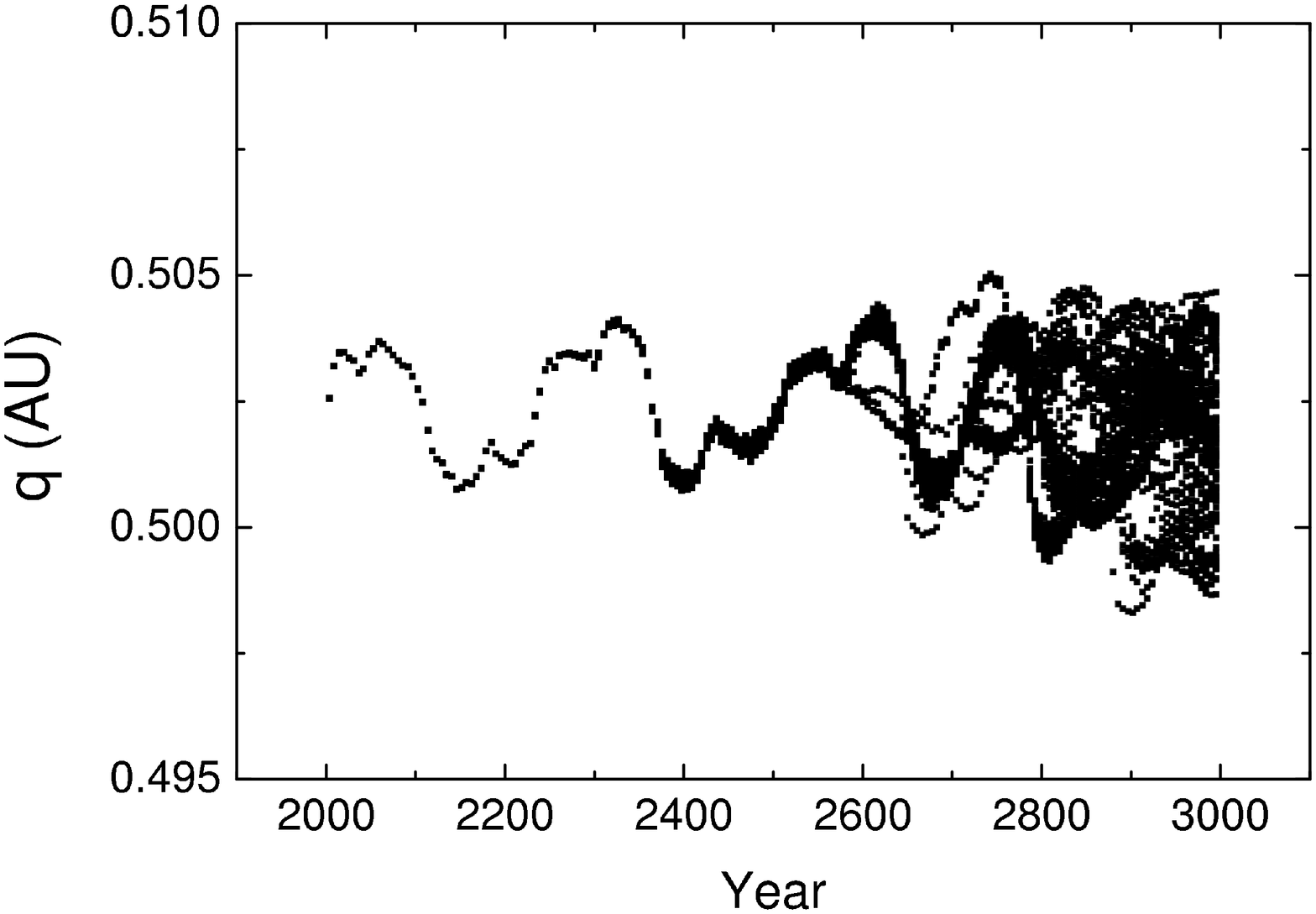}
            \hspace{0.6cm}
            \includegraphics[width=3.4cm,angle=-90]{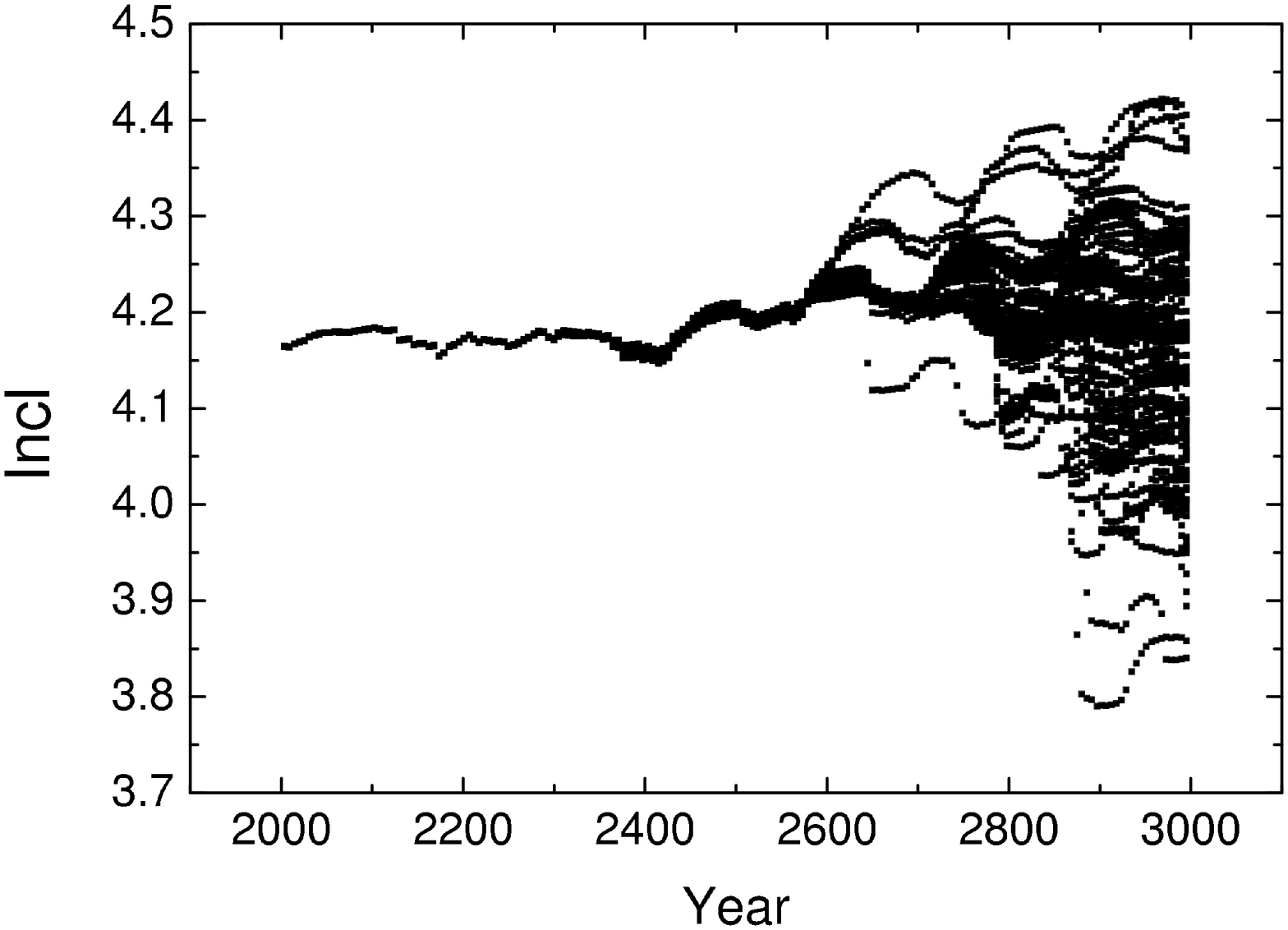}}
\vspace{0.4cm} \centerline{\includegraphics[width=3.4cm,angle=-90]{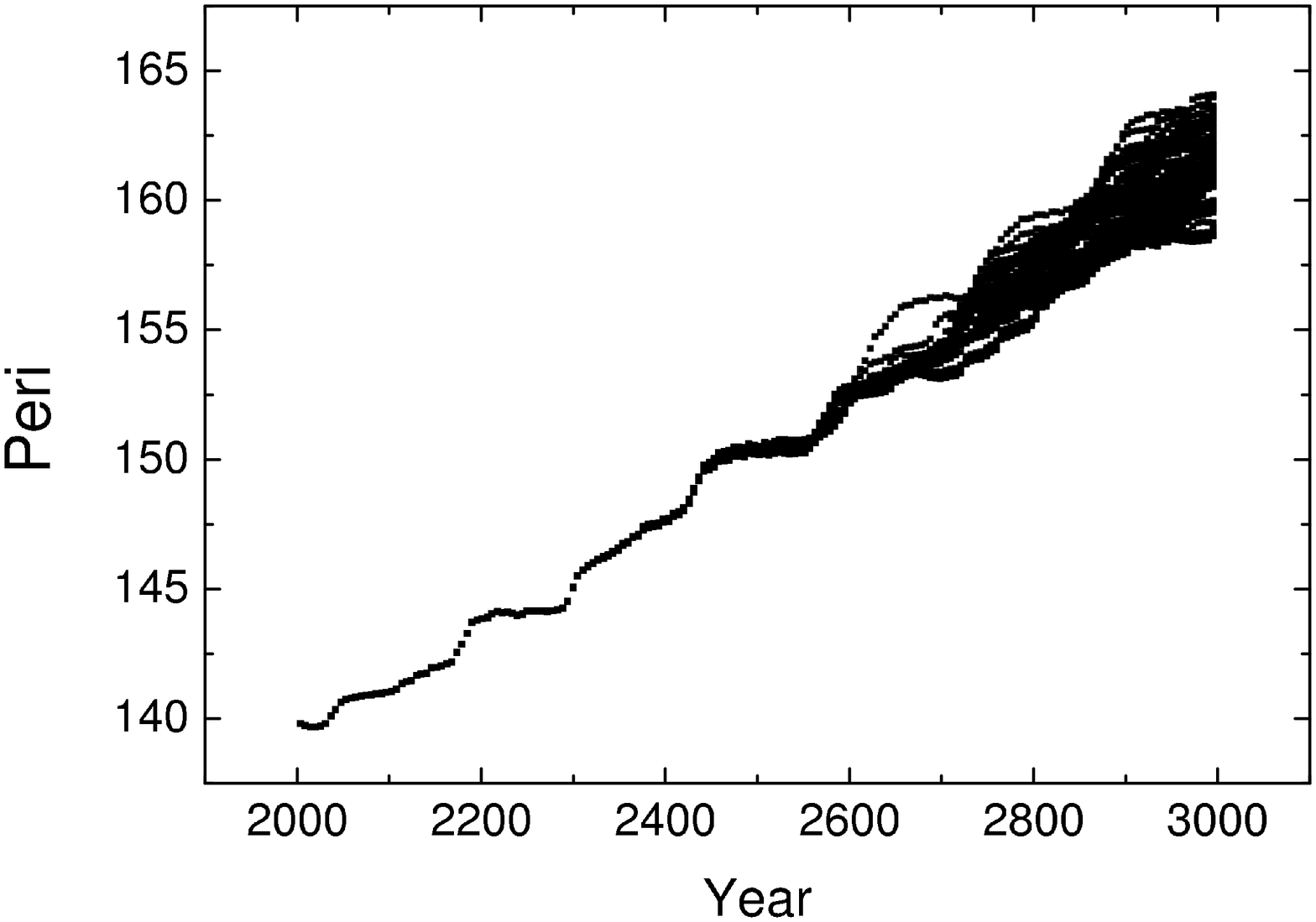}
            \hspace{0.6cm}
            \includegraphics[width=3.4cm,angle=-90]{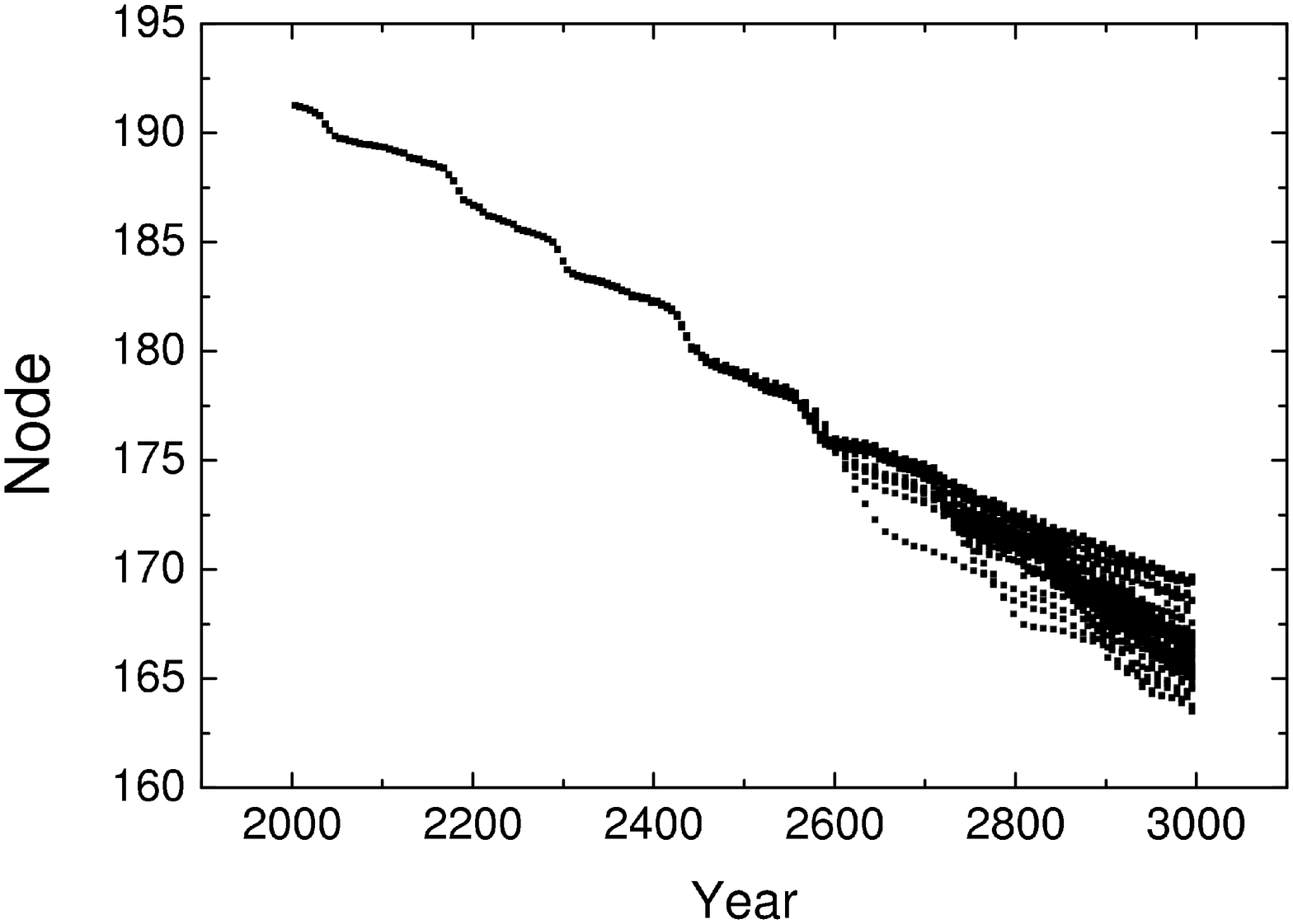}}
\vspace{0.4cm} \centerline{\includegraphics[width=3.4cm,angle=-90]{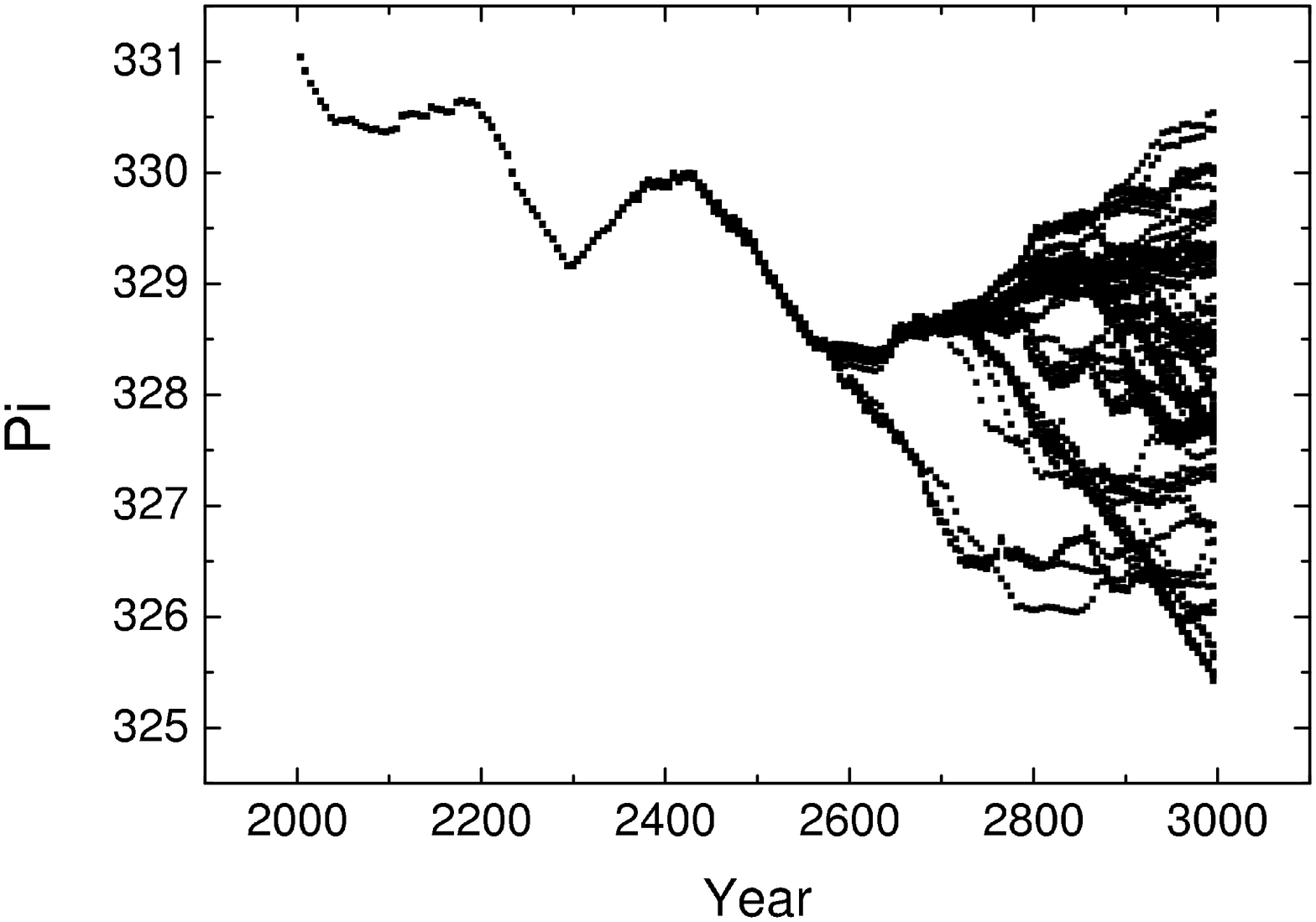}
            \hspace{0.6cm}
            \includegraphics[width=3.4cm,angle=-90]{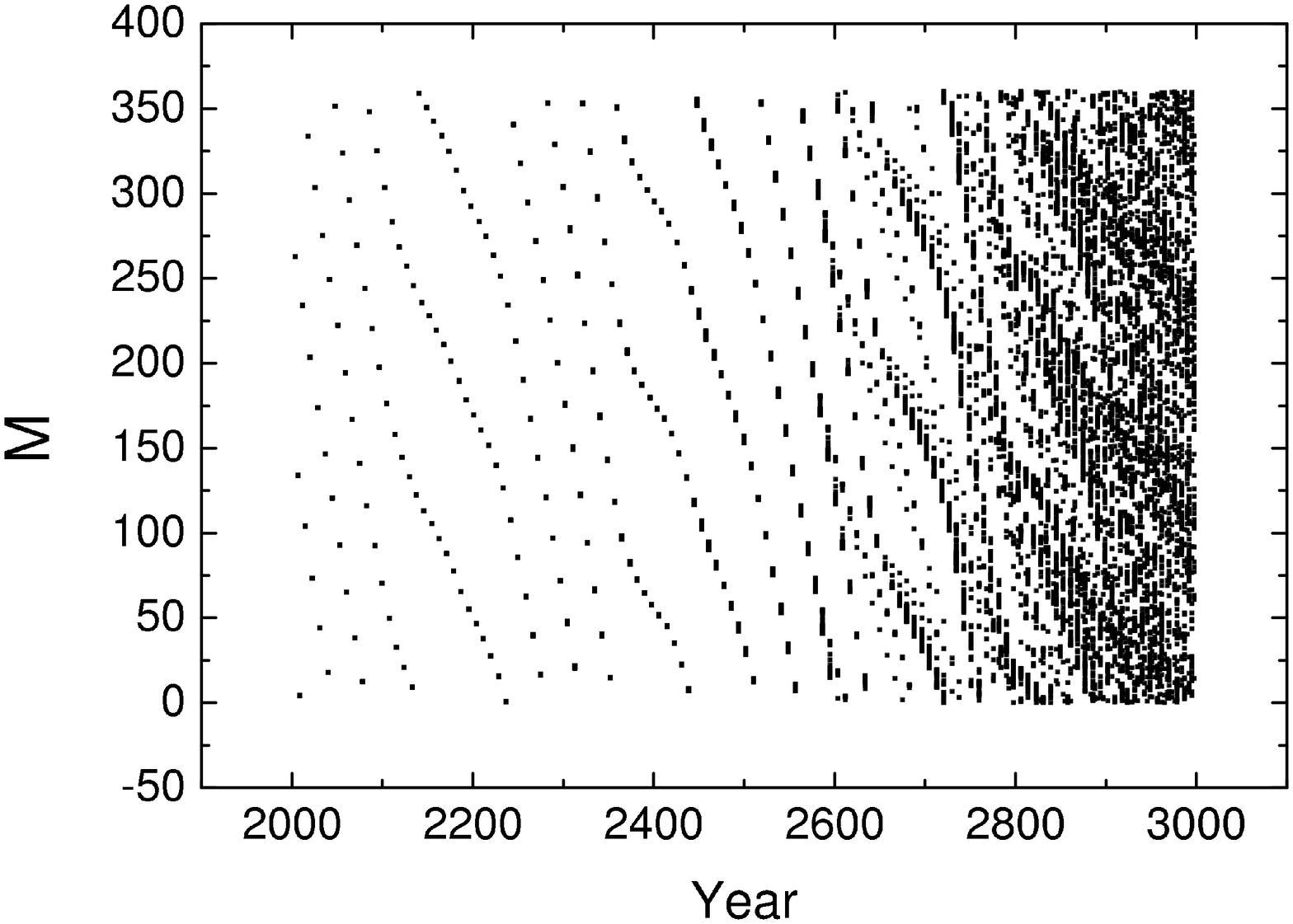}}
\caption{The orbital evolution of modeled particles released by Earth's tides
over 1000 years; $a$ - semimajor axis, $ecc$ - eccentricity, $q$ - perihelion
distance, $Incl$ - inclination, $Peri$ - argument of perihelion, $Node$ -
longitude of the ascending node, $Pi$ - longitude of perihelion, $M$ - mean
anomaly.}
\end{figure}

During the approach to the Earth the model particles start to leave the surface
of the parent asteroid but fall back after a short time. We define the Roche
limit for our purpose ($RL_a$) as a distance from the center of the Earth such
that the particle did not fall back onto the asteroid surface. For 2004\,FU162
we found $RL_a=22\,280$\,km and the asteroid was traveling 50 minutes inside
this limit. The P\v{r}\'{i}bram-like asteroid travels 13 minutes inside its
$RL_a=14\,700$\,km. Obviously, the fly-by time depends on the approach geometry
and the relative velocity of the asteroid to the Earth. For our model orbits we
set the approach epochs to the real approach time of 2004\,FU162 and fall time
of P\v{r}\'{i}bram, April 7, 1959. Adequate velocities in perigee were
13.4\,km\,s$^{-1}$ for 2004 FU162 and 19.2\,km\,s$^{-1}$ for P\v{r}\'{i}bram,
respectively.

Particles that definitely left 2004\,FU162 surface reached escape velocities in
range of 5-10 \,cm\,s$^{-1}$ and in the geocentric distance of asteroid
100\,000\,km leaving particles were $\sim$ hundreds of meters away from the
parent body (Table 1). We also considered a possible mass loss of the parent
body that could theoretically change the behavior of leaving particle motion.
Though the model took two possibilities into account (after the perigee the
whole parent asteroid survives or the half of the mass survives), the parent
body mass loss had minimum influence on particle motion and particle cloud
diffusion. The particle gains the maximum escape velocity when leaving the
parent body before the perigee.

%Fig. 2
\begin {figure}
\centerline{\includegraphics[width=3.4cm,angle=-90]{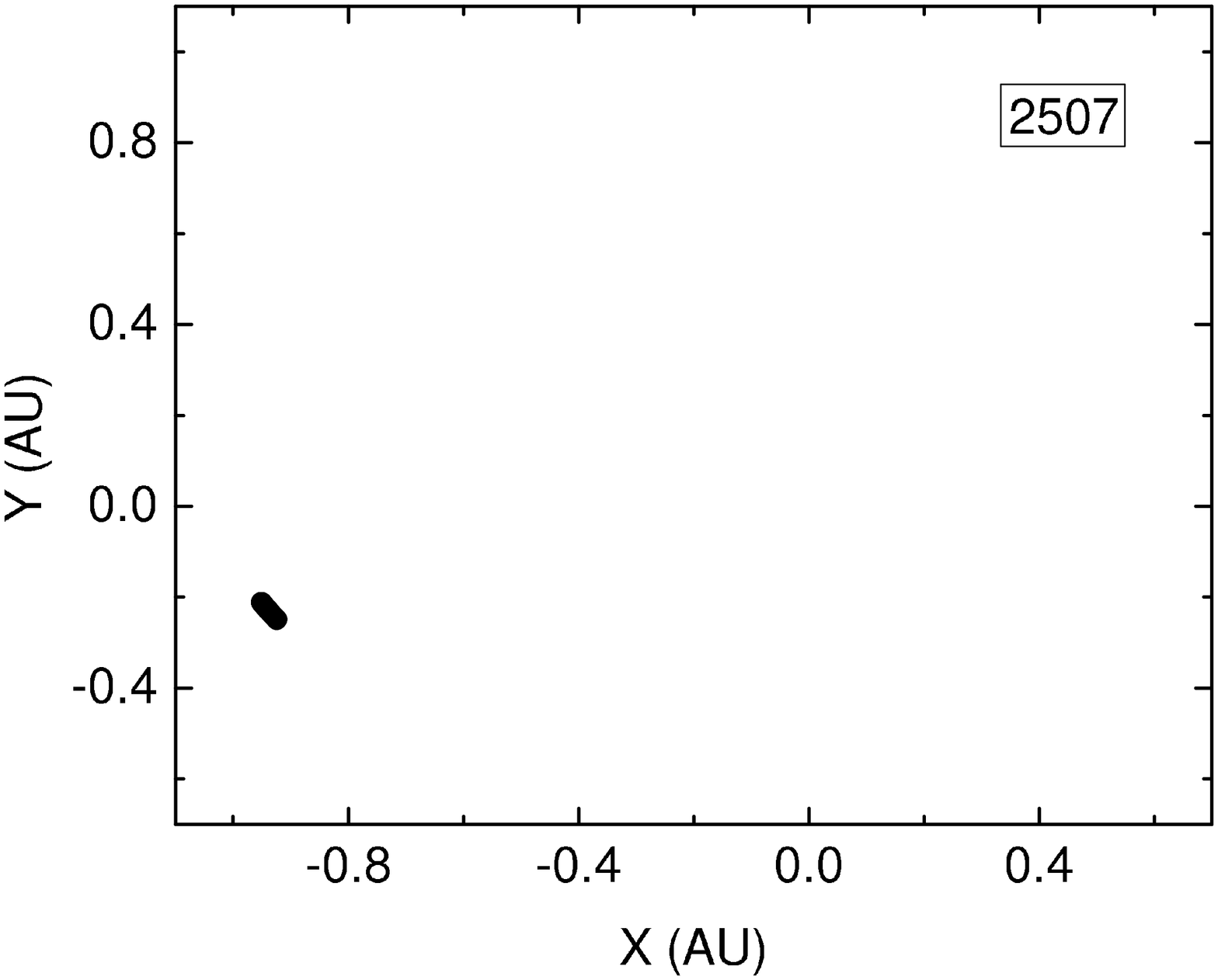}
            \hspace{0.6cm}
            \includegraphics[width=3.4cm,angle=-90]{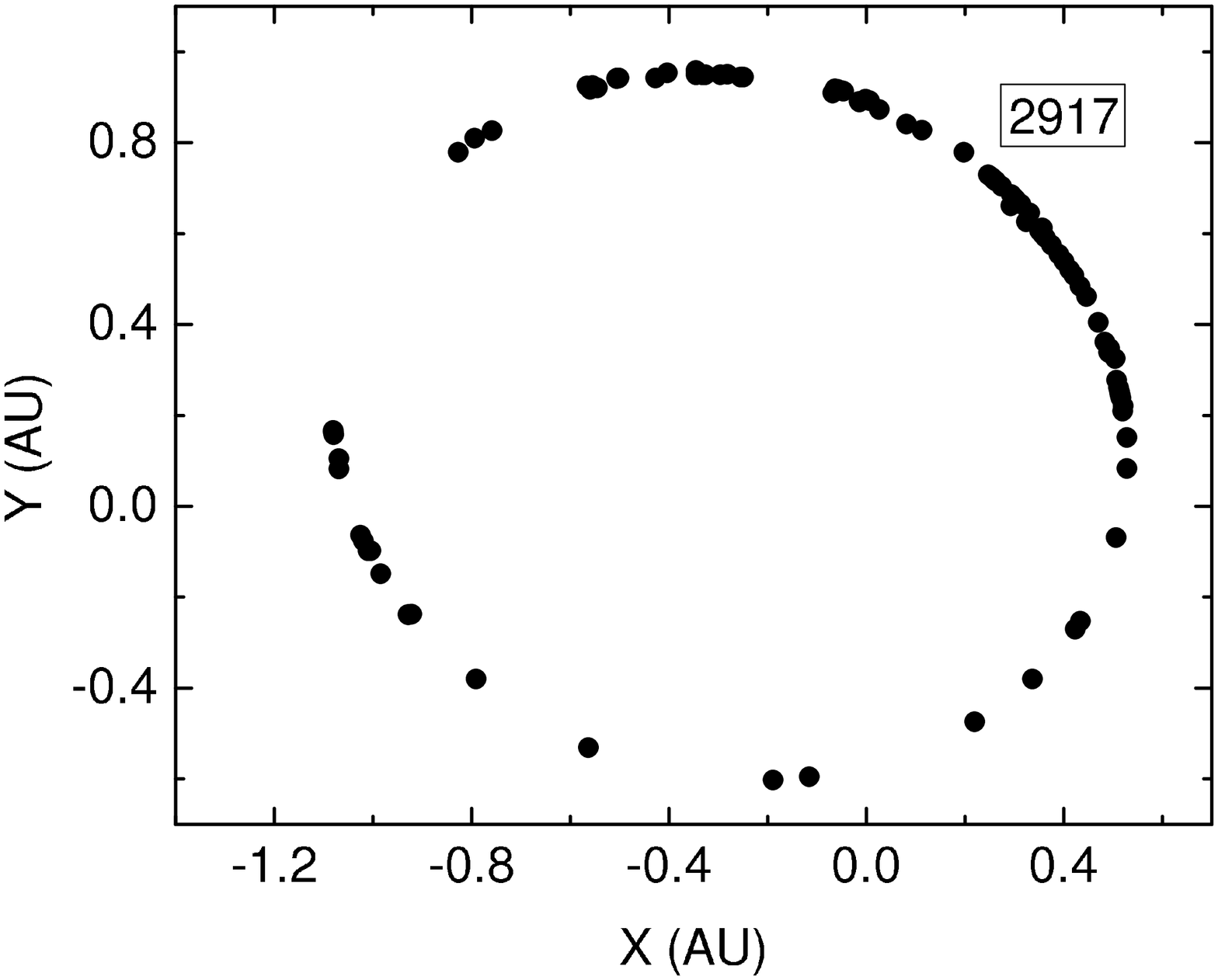}}
\caption{The spread of modeled particles in years 2507 and 2917, respectively.
Projection onto the ecliptic plane ($x,y$).}
\end{figure}

The motion of a cloud of 100 escaping particles was examined through its motion
during next 1000 years. In the first 500 years the orbital evolution of all
particles is practically identical with no significant spread. Figure\,1 shows
the evolution of orbital elements of particles that left the parent body of
half the mass of Itokawa on a 2004\,FU162-like orbit. In this case the
particles obtained maximum escape velocities and their spread in orbital
elements at the end of integration was widest as well. The nature of orbital
elements evolution of particles derived from a P\v{r}\'{i}bram-like orbit is
similar, but with smaller spread.

Figure\,2 shows the cloud of test particles projected onto the ecliptic plane.
The spread of the particles demonstrates that during first few hundred years
the particles remain in one packet and later disperse around the orbit of the
parent body.

\section{Conclusions}

The gravitational tidal influence of the Earth on potential meteoroids
originating from approaching asteroids is very small even during the close
approach. On the other hand, the escape velocity of a pebble, rock or a boulder
from a small asteroid is only several cm\,s$^{-1}$. We found that the escape
velocity of a particle lying on the asteroid surface and its possible
independent motion depends on the geometry of the parent body fly-by of the
Earth. Although the different geocentric velocity allows the approaching
asteroids to stay longer or shorter inside the Roche limit where the particle
release is possible, our simulation showed that particles left from two
different geocentric orbits obtained very similar escape velocities. That is
why we suppose that the meteoroid streams from parent bodies on different
orbits will behave similar in order of stream dispersion and evolution. Even a
simple model of meteoroid particles production by tidal disruption of asteroids
shows that the orbital evolution of particles is stable at least for the first
1000 years. Released particles stay in a relative small cloud for about 500
years and spread evenly along the orbit of the parent asteroid during next
several hundred years. Their orbital elements exhibit a very small dispersion
in the mentioned time frame.

Accordingly, we expect similar meteoroid stream behavior of a possible
asteroidal origin. If there is observed such kind of the stream, our model
might be useful for interpretation of asteroidal stream origin.

\acknowledgements This work was supported by VEGA - the Slovak Grant Agency for
Science, grant No. 1/3067/06, and by Comenius University Grant No. UK/399/2008.

\end{document}